\documentclass[a4paper,fleqn]{cas-dc}
\usepackage{amsmath,amsfonts}
\usepackage{algorithmic}
\usepackage{algorithm}
\usepackage{array}
\newcolumntype{P}[1]{>{\centering\arraybackslash}p{#1}}
\newcolumntype{M}[1]{>{\centering\arraybackslash}m{#1}}
\usepackage{subfig}
\usepackage{textcomp}
\usepackage{stfloats}
\usepackage{url}
\usepackage{verbatim}
\usepackage{graphicx}
\usepackage{multicol}
\usepackage{multirow}
\usepackage[capitalise]{cleveref} 
\usepackage{booktabs}

\usepackage{enumitem}
\usepackage{longtable}
\usepackage{framed}
\usepackage{mdframed}
\usepackage{tcolorbox}
\usepackage[flushleft]{threeparttable}
\usepackage{nomencl}
\makenomenclature
\renewcommand*\nompreamble{\begin{multicols}{2}}
\renewcommand*\nompostamble{\end{multicols}} 

\usepackage[numbers]{natbib}

\def\tsc#1{\csdef{#1}{\textsc{\lowercase{#1}}\xspace}}
\tsc{WGM}
\tsc{QE}

\begin{document}
\let\WriteBookmarks\relax
\def\floatpagepagefraction{1}
\def\textpagefraction{.001}


\shortauthors{Thien-Nu Hoang}  

\title [mode = title]{Detecting In-vehicle Intrusion via Semi-supervised Learning-based Convolutional Adversarial Autoencoders}  



%

\author[1]{Thien-Nu Hoang}[]






\affiliation[1]{organization={Department
of Future Convergence Technology, Soonchunhyang University},
            addressline={}, 
            city={Asan-si},
            postcode={31538}, 
            state={Chuncheongnam-do},
            country={South Korea}}

\author[1]{Daehee Kim}[]
\cormark[1]

\ead{daheekim@sch.ac.kr}




\cortext[1]{Corresponding author}



\begin{abstract}
With the development of autonomous vehicle technology, the controller area network (CAN) bus has become the de facto standard for an in-vehicle communication system because of its simplicity and efficiency. However, without any encryption and authentication mechanisms, the in-vehicle network using the CAN protocol is susceptible to a wide range of attacks. Many studies, which are mostly based on machine learning, have proposed installing an intrusion detection system (IDS) for anomaly detection in the CAN bus system. Although machine learning methods have many advantages for IDS, previous models usually require a large amount of labeled data, which results in high time and labor costs. To handle this problem, we propose a novel semi-supervised learning-based convolutional adversarial autoencoder model in this paper. The proposed model combines two popular deep learning models: autoencoder and generative adversarial networks. First, the model is trained with unlabeled data to learn the manifolds of normal and attack patterns. Then, only a small number of labeled samples are used in supervised training. The proposed model can detect various kinds of message injection attacks, such as DoS, fuzzy, and spoofing, as well as unknown attacks. 
The experimental results show that the proposed model achieves the highest F1 score of 0.99 and a low error rate of 0.1\% with limited labeled data compared to other supervised methods. In addition, we show that the model can meet the real-time requirement by analyzing the model complexity in terms of the number of trainable parameters and inference time. This study successfully reduced the number of model parameters by five times and the inference time by eight times, compared to a state-of-the-art model.
\end{abstract}



\begin{highlights}
\item Propose a novel semi-supervised learning model for the in-vehicle intrusion detection system named convolutional adversarial autoencoders.
\item Detect both known and unknown attacks effectively with a small number of labeled samples.
\item Reduce the model complexity significantly in terms of the trainable parameters and the inference time which are essential for the real-time in-vehicle intrusion detection.
\end{highlights}


\maketitle

\mbox{}

\nomenclature{CAN}{Controller Area Network}
\nomenclature{ECU}{Electrical Control Units}
\nomenclature{CAAE}{Convolutional Adversarial Autoencoders}
\nomenclature{AAE}{Adversarial Autoencoders}
\nomenclature{IDS}{Intrusion Detection System}
\nomenclature{GAN}{Generative Adversarial Networks}
\nomenclature{DCNN}{Deep Convolutional Neural Network}
\nomenclature{LSTM}{Long Short-Term Memory}
\nomenclature{GRU}{Gated Recurrent Unit}
\nomenclature{SVM}{Support Vector Machine}
\nomenclature{DT}{Decision Trees}
\nomenclature{RF}{Random Forest}
\nomenclature{MLP}{Multi-Layer Perceptron}
\nomenclature{KNN}{K-Nearest Neighbors}
\nomenclature{OSVM}{One-class Support Vector Machine}
\nomenclature{HTM}{Hierarchical Temporal Memory}
\nomenclature{SGD}{Stochastic Gradient Descent}
\nomenclature{ANN}{Artificial Neural Network}
\nomenclature{HEX}{Hexadecimal Representation}
\nomenclature{GP}{Gradient Penalty}
\nomenclature{DAE}{Deep AutoEncoder}


\begin{table*}[t]
\begin{mdframed}
 \printnomenclature   
\end{mdframed}
\end{table*}

\section{Introduction}

\begin{figure*}[ht!]
\centering
\includegraphics[width=5in]{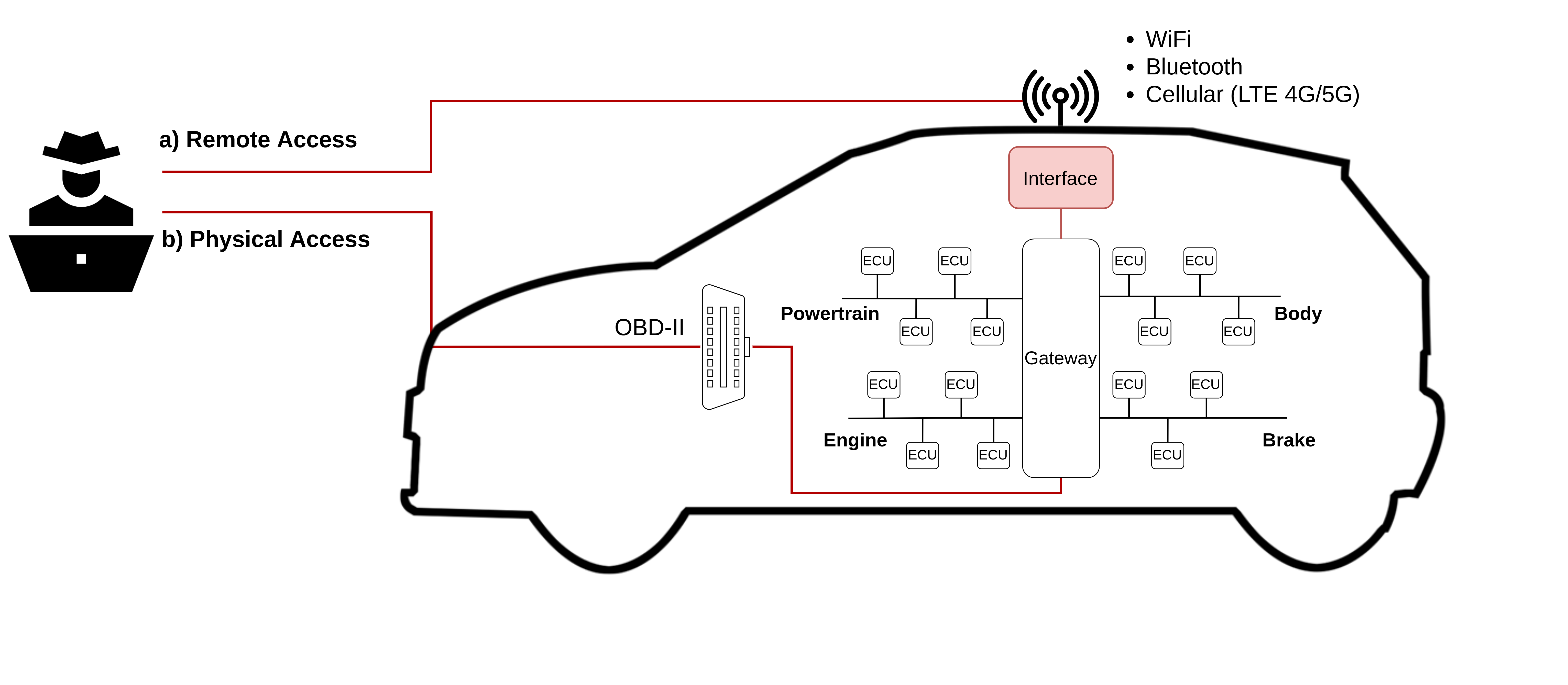}   
\caption{Overview of an in-vehicle network using a CAN bus. Many buses installed in the vehicle include multiple ECUs to serve a specific function and are connected through a gateway. The network can be accessed through an OBD-II port and wireless technologies such as WiFi, Bluetooth, and cellular network.}
\label{fig:can_bus_sys}
\end{figure*}

%
%
%
%




To enable autonomous vehicles, many sensors or electrical control units (ECUs) have been installed on vehicles \cite{Tuohy2015Intra-VehicleReview}. The ECUs provide useful information about the real-life environment to the vehicles, which facilitates the learning process of the vehicles through deep learning \cite{Deo2018Multi-ModalLSTMs, Lee2021FastTransformation} or reinforcement learning methods \cite{Isele2017NavigatingLearning, Wang2018AManeuvers}. In addition, these connected ECUs communicate with each other through a controller area network (CAN) bus system \cite{1991BOSCH2.0}. Since 1986, the CAN bus system has been used widely because of its high speed and efficiency. Furthermore, the CAN bus operates based on broadcast and message priority mechanisms. However, there is no authentication and encryption in the CAN bus. Therefore, the system is vulnerable to various cyberattacks, particularly message injection attacks. It can be risky when an attacker accesses the CAN bus system and sends anomalous messages. For example, the attacker can control the brake, lock the doors, or even steer the vehicle. Many studies have shown that we can inject messages to the CAN bus system directly through an on-board diagnostics II (OBD-II) port or indirectly through WiFi or bluetooth channels \cite{Koscher2010ExperimentalAutomobile, Hoppe2011SecurityCountermeasures, Jo2021ACountermeasures}.  

An intrusion detection system (IDS) has been introduced to monitor and detect attacks in CAN buses \cite{Hoppe2011SecurityCountermeasures}. IDS, which is usually placed in an ECU, receives and analyses incoming messages. It will raise an alert to other ECUs if any anomalous message is detected. The machine learning-based IDS has piqued the interest of many researchers because it can adapt to the complexity and instability of cybersecurity in the in-vehicle network. In terms of detecting manner, the machine learning-based IDS can be divided into two groups: unsupervised and supervised models. On the one hand, unsupervised models learn the representation of normal patterns and then detect an abnormal state based on its deviation from the normal. The problem with this scheme is that the model produces a high false-positive rate. In addition, we need to find an optimal threshold to achieve the best result. On the other hand, supervised models learn to classify a predefined number of classes provided by a labeled dataset. Therefore, we usually must provide a large amount of labeled data to supervised models to achieve a good outcome. 




In this study, we propose a novel semi-supervised deep learning-based IDS, in which the model learns to extract appropriate features from unlabeled data by itself before being trained in a supervised manner. Therefore, the proposed model can handle not only limited data environments but also unknown attacks.
Specifically, our main contributions can be summarized as follows:
\begin{itemize}
    \item We propose a convolutional adversarial autoencoder (CAAE) based IDS by combining convolutional autoencoder and generative adversarial networks (GAN) to counter both known and unknown attacks. Because the proposed model is trained in semi-supervised learning, only a small number of labeled data is required during training. We believe that the proposed model can reduce the time to collect and annotate data significantly. To the best of our knowledge, this is the first time that CAAE is applied to the in-vehicle IDS.
    \item To demonstrate the performance of our model, we conducted a comprehensive experiment using a real-car dataset 
    with both known and unknown attacks. In addition, we provide the source code\footnote{Source code is available at https://github.com/htn274/CanBus-IDS} to facilitate future studies on this topic.
    \item Using approximately 60k labeled samples, which accounts for only 40\% of the total training data, the proposed method achieved a high F1 score of 0.99 and a low error rate of 0.1\%, compared to other supervised and unsupervised models. Moreover, the proposed model successfully reduced the number of model parameters by 5 times and 8 times for the inference time. Therefore, the proposed model is efficient for real-time detection.
\end{itemize}

The remaining part of the paper proceeds as follows: Section \ref{sec:background} introduces the background of the CAN bus system and attack models related to the study. Related works are presented in Section \ref{sec:related_works}. Furthermore, Section \ref{sec:method} describes our proposed method in detail. The experimental results and conclusion are described in Section \ref{sec:results} and Section \ref{sec:conclusion}, respectively.

\section{Background and attack model \label{sec:background}}

\begin{figure*}[t]
\centering
\includegraphics[width=6in]{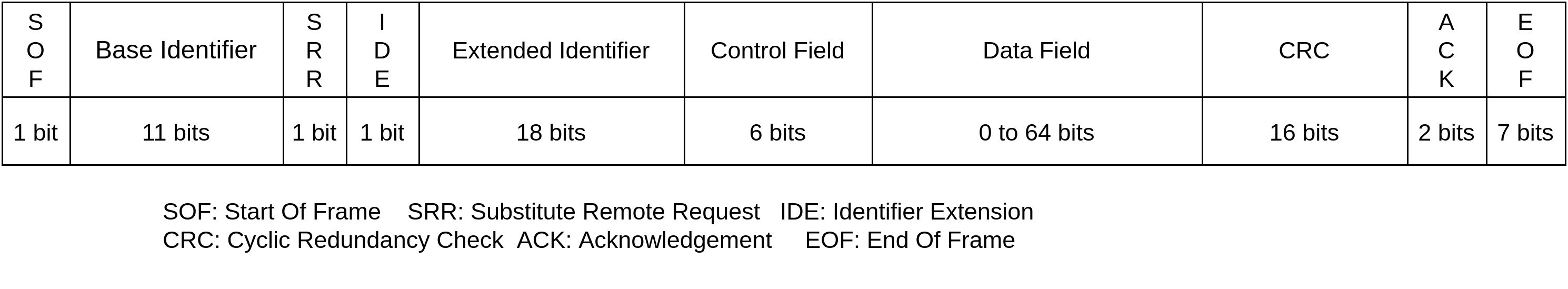}   
\caption{CAN2.0B data frame format.}
\label{fig:can_data_frame}
\end{figure*}

\begin{figure}[t]
\centering
\subfloat[DoS attack \label{fig:dos}]{{\includegraphics[width=0.25\textwidth]{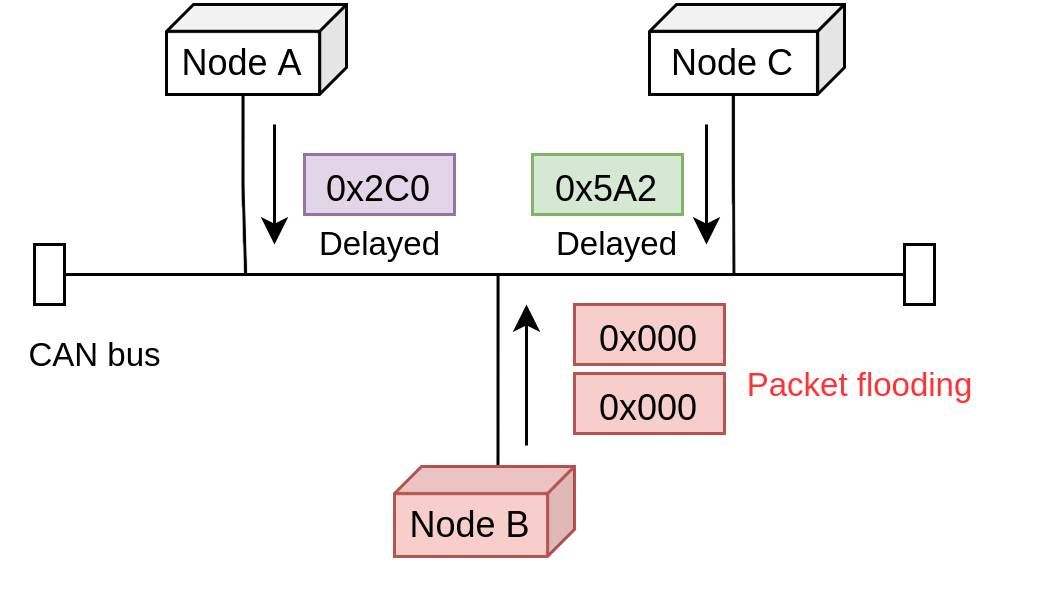}}}\hfill
\subfloat[Spoofing attack \label{fig:spoof}]{{\includegraphics[width=0.25\textwidth]{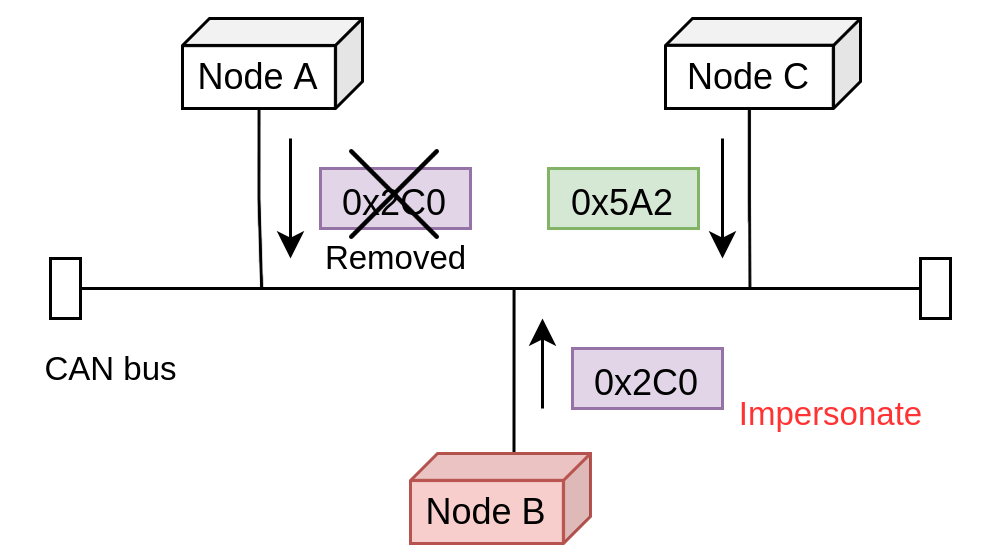}}}
\subfloat[Fuzzy attack \label{fig:fuzzy}]{{\includegraphics[width=0.25\textwidth]{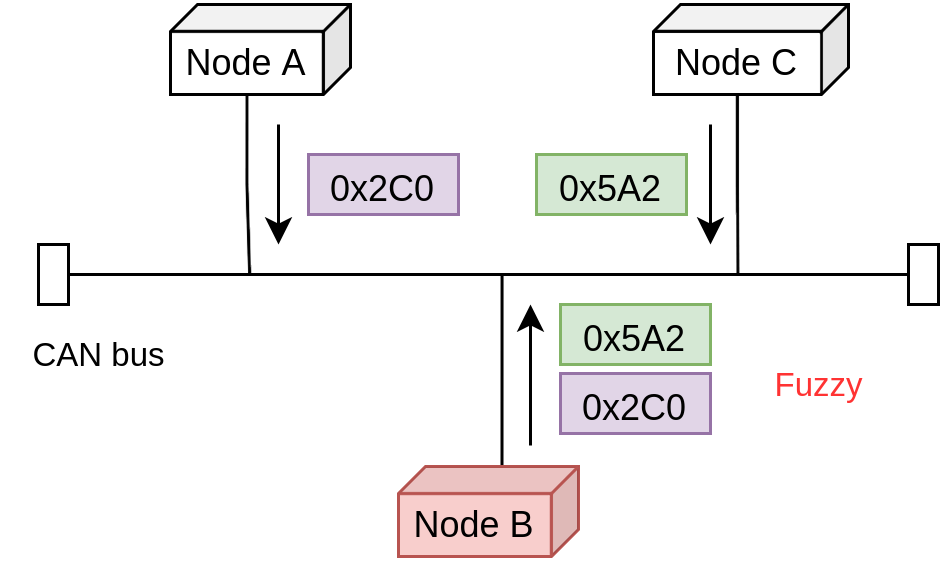}}}
\caption{Depiction of three attack models.}
\label{fig:attack_models}
\end{figure}

\subsection{CAN bus system}

The CAN bus system \cite{1991BOSCH2.0}, which was first introduced by Robert Bosh GmbH in 1985, is a channel for communication between ECUs. Because the CAN bus reduces the complexity and cost of in-vehicle wiring, it has become a de-facto standard for in-vehicle communication systems. A modern vehicle can comprise multiple CAN buses. Each bus supports a specific function (e.g., engine control, powertrain control, brake control, and others \cite{Yoshiyasu2018VehicleSecurity}). These buses are connected through a gateway, as illustrated in \cref{fig:can_bus_sys}.

ECUs exchange information with others through the CAN bus system by broadcasting messages following a predefined data frame format. Each type of message is identified by a CAN ID, which describes the meaning of the data. Therefore, the CAN ID helps receiver ECUs choose appropriate messages for specific functions. In addition, ECUs can be added to the vehicle without any modification to the CAN messages, rendering the CAN bus system more flexible. According to the length of the CAN ID, CAN messages are classified into two types: CAN 2.0A and CAN 2.0B. The ID of CAN 2.0A is only 11 bits (including the base identifier), whereas the ID of CAN 2.0B is 29 bits (including both base and extended identifiers) (see \cref{fig:can_data_frame}). In addition, the CAN ID is used to determine the priority of multiple messages. For example, a message with more leading zero bits in its ID wins the right to be transmitted. In addition, there is a 64-bit data field that contains the information that the sender ECUs want to transmit. The other fields are not involved in IDS research and thus are not explained here. We refer to the CAN specification \cite{1991BOSCH2.0} for additional information on how the CAN bus system works.

\subsection{Attack model}

Because of the working principle of the CAN bus system: broadcast-based system and message priority scheme without encryption and authentication mechanisms, the system is vulnerable to cyber-attacks. Before launching an actual attack, an attacker must access the CAN-bus system. This assumption is practical as many studies launched attacks on the in-vehicle network successfully both directly (via an OBD-II port) and indirectly (via wireless technologies such as WiFi, cellular network, and Bluetooth) \cite{Koscher2010ExperimentalAutomobile, Hoppe2011SecurityCountermeasures, Jo2021ACountermeasures}. After the network is accessed, the attacker can inject malicious messages, resulting in various kinds of attacks such as DoS, spoofing, and fuzzy, which are described in detail below and illustrated in \cref{fig:attack_models}: 

\begin{itemize}
    \item DoS: The DoS attack aims to disable the availability of the network. By utilizing the message priority mechanism, the attacker can inject high-priority messages to win the bus and prevent transmission of other legitimate messages. For example, the attacker (node B) in \cref{fig:dos} injects the highest priority message whose ID is all zero bits (0x000). As a result, legitimate messages from other nodes (A and C) are delayed. 
    \item Spoofing: The spoofing attack aims to impersonate a CAN ID to control a specific function of the vehicle as the attacker desires. To accomplish that, the attacker must first determine the target CAN ID and then inject messages with that ID. For example, the attacker in \cref{fig:spoof} masquerades CAN ID of (0x2C0) and sends manipulated messages because there is no authentication.
    \item Fuzzy: The fuzzy attack aims to make the vehicle malfunction as well as delay other messages. The attacker can inject multiple messages with a random CAN ID and data. For example, the attacker in \cref{fig:fuzzy} sends malicious messages with random CAN IDs, causing the vehicle to malfunction. 
\end{itemize}

\section{Related works \label{sec:related_works}}

An IDS monitors and detects malicious behaviors in a network. For the in-vehicle network, an IDS can be installed in an ECU to serve as an additional node in the CAN bus. Because each ECU broadcasts messages, the IDS analyzes incoming messages and alerts if any abnormality exists in the network. In terms of design, the in-vehicle IDS can be grouped into four categories: fingerprint-based methods (bus level), parameter monitoring-based methods (message level), information theoretic-based methods (data-flow level), and machine learning-based methods (functional level) \cite{Wu2020ANetworks}. Previous studies showed that machine learning methods are efficient for detecting an attack from the application layer. Our study also falls into this category. In this section, we will review state-of-the-art machine learning studies related to in-vehicle IDS, which are summarized in \cref{tab:litrev}. 

For intruder detection problems, machine learning or deep learning models can be trained in supervised or unsupervised manners. Regarding supervised approaches, a large and accurate labeled dataset is required. The IDS problem can be formulated as a binary classification with two classes: normal and abnormal. In \cite{Kang2016ASecurity}, the authors proposed a deep neural network consisting of 64 neurons at the input layer, which represents the data field in the CAN message. They tested the proposed model with a simulation dataset and achieved a high detection rate for both the classes. By contrast, the authors from \cite{Song2020In-vehicleNetwork} published a real dataset - the car hacking dataset, which has been widely used to test IDS models. They also proposed a deep convolutional neural network (DCNN) that accepts a 29-CAN-ID frame as input. Although the DCNN model has a low false alarm rate, it requires high computation costs. Furthermore, the authors in \cite{Hossain2020LSTM-basedCommunications} leverage the time-series information for IDS, using long short-term memory (LSTM) neural networks. In addition, a combination of CNN and attention-based gated recurrent unit (GRU) was proposed in \cite{Javed2021CANintelliIDS:GRU}. Similar to DCNN, the proposed models are extremely complicated to deploy in real life. Conversely, in \cite{Moulahi2021ComparativeBus}, simple machine learning models are used for faster training and inference. However, the models achieve low accuracy, particularly for DoS and fuzzy attacks.





Although the supervised models provide good results, it is difficult to collect sufficient labeled data for learning. In addition, some models cannot detect zero-day attacks because the supervised models can only learn existing patterns in the training dataset. Consequently, unsupervised models have been proposed, in which only normal samples are used in the training phase. In the testing phase, any sample deviating very far from the learned patterns is classified as abnormal. Based on this concept, some in-vehicle IDS studies proposed traditional machine learning techniques, such as K-means and K-nearest neighbors \cite{DAngelo2020AVehicles}, one-class SVM (OSVM) \cite{Avatefipour2019AnLearning}, and Kohonen SOM network (KSOM) \cite{Barletta2020IntrusionApproach}, whereas others proposed deep learning models, such as hierarchical temporal memory (HTM) learning algorithm \cite{Wang2018AHTM} and LSTM based autoencoder \cite{Ashraf2020NovelSystems} to improve the detection performance. However, the unsupervised models perform worse than the supervised models because of the high false-positive rate. 
 




To fill the gap between supervised and unsupervised IDS, the authors in \cite{Seo2018GIDS:Network} proposed a two-stage deep neural network: the first classifier is trained in a supervised manner, whereas the second one is a discriminator in a GAN network and is used for detecting unknown attacks. They evaluated the two classifiers separately, and the combined result was not reported. A new idea presented in \cite{Song2021Self-SupervisedData} is to generate attack samples by an LSTM-based GAN model, and then the generated samples and available normal samples are fed into a DCNN model. The study is promising but achieved low accuracy and needs to be further developed. The authors from \cite{Yang2021MTH-IDS:Vehicles} used tree-based machine learning algorithms and focused on developing a complicated data preprocessing framework to improve the accuracy. 

Compared to existing studies related to in-vehicle IDS, our proposed model has some advantages as follows: 1) It is trained end-to-end using a small number of labeled data without any complicated data preprocessing; 2) It can detect both known and unknown attacks with high precision and recall, compared to other models; 3) It processes a new sample within a millisecond, which meets the real-time requirement for the in-vehicle IDS.

\begin{table*}[t]
    \centering
    \setlength{\leftmargini}{0.4cm}
    \caption{Literature reviews of machine-learning-based IDS for the in-vehicle network.}
    \begin{tabular}{|M{1.7cm}|M{1.5cm}|M{2cm}|M{2.5cm}|M{3cm}|M{3cm}|}
         \hline \textbf{Categories} & \textbf{Research Work} &\textbf{ ML algorithm} & \textbf{Features} & \textbf{Contributions} & \textbf{Limitations}  \\
         \hline 
         \multirow{4}{*}{Supervised} & \cite{Kang2016ASecurity} &
        \begin{minipage}[t][1.3cm]{\linewidth}
          \begin{itemize}
             \item DNN
         \end{itemize}           
        \end{minipage} 
          & 
        \begin{minipage}[t]{\linewidth}
          \begin{itemize}
             \item Data payload
         \end{itemize} 
         \end{minipage}
         & 
        \begin{minipage}[t]{\linewidth}
         \begin{itemize}
             \item Lightweight and \newline fast model
         \end{itemize}  
         \end{minipage} & 
        \begin{minipage}[t]{\linewidth}
         \begin{itemize}
             \item Train and test on a \newline simuluation dataset
         \end{itemize} 
         \end{minipage} \\  
         \cline{2-6} & \cite{Song2020In-vehicleNetwork} & 
        \begin{minipage}[t][1.1cm]{\linewidth}
         \begin{itemize}
             \item DCNN
         \end{itemize} 
        \end{minipage} 
         & \begin{minipage}[t]{\linewidth}
         \begin{itemize}\item CAN IDs \end{itemize}
        \end{minipage}  & \begin{minipage}[t][1.3cm]{\linewidth}
         \begin{itemize}
             \item Novel data \newline processing technique
         \end{itemize}
        \end{minipage} & \begin{minipage}[t]{\linewidth}
         \begin{itemize}
             \item Complex model
         \end{itemize}
        \end{minipage}  \\ 
         \cline{2-6} & \cite{Hossain2020LSTM-basedCommunications} & \begin{minipage}[t][1.3cm]{\linewidth}
         \begin{itemize}
             \item LSTM
         \end{itemize}
        \end{minipage}  & \begin{minipage}[t][1.1cm]{\linewidth}
         \begin{itemize}
             \item CAN IDs
             \item Data payload
         \end{itemize}
        \end{minipage}  & \begin{minipage}[t][1.1cm]{\linewidth}
         \begin{itemize}
             \item Best tuned \newline parameters for \newline LSTM-based IDS
         \end{itemize}
        \end{minipage}  & \begin{minipage}[t][1.1cm]{\linewidth}
         \begin{itemize}
             \item Need a large \newline labeled training \newline dataset
         \end{itemize}
        \end{minipage}   \\
         \cline{2-6} & \cite{Javed2021CANintelliIDS:GRU} &
         \begin{minipage}[t][1.5cm]{\linewidth}
         \begin{itemize}
             \item 1-D CNN
             \item Attention-based GRU
         \end{itemize}
        \end{minipage} 
           & \begin{minipage}[t][1.1cm]{\linewidth}
         \begin{itemize}
             \item Time stamp
             \item CAN IDs
             \item Data payload
         \end{itemize}
        \end{minipage}  & \begin{minipage}[t][1.1cm]{\linewidth}
         \begin{itemize}
             \item Novel way for \newline features extraction
         \end{itemize}
        \end{minipage}  & \begin{minipage}[t][1.1cm]{\linewidth}
         \begin{itemize}
             \item Complex model 
             \item No real-time \newline evaluation
         \end{itemize}
        \end{minipage}  \\ 
         \cline{2-6} & \cite{Moulahi2021ComparativeBus} &\begin{minipage}[t][2cm]{\linewidth}
         \begin{itemize}
             \item SVM
             \item DT
             \item RF
             \item MLP
         \end{itemize}
        \end{minipage} &\begin{minipage}[t][1.1cm]{\linewidth}
         \begin{itemize}
             \item Time stamp
             \item CAN IDs
             \item Data payload
         \end{itemize}
        \end{minipage}  & \begin{minipage}[t][1.1cm]{\linewidth}
         \begin{itemize}
             \item Short training time
         \end{itemize}
        \end{minipage}  & \begin{minipage}[t][1.1cm]{\linewidth}
         \begin{itemize}
             \item Inefficient for DoS and fuzzy attacks
         \end{itemize}
        \end{minipage}  \\
         \hline 
         \multirow{4}{*}{Unsupervised} & \cite{DAngelo2020AVehicles} & \begin{minipage}[t][1.8cm]{\linewidth}
         \begin{itemize}
             \item K-means
             \item KNN
         \end{itemize}
        \end{minipage}  & \begin{minipage}[t][1.8cm]{\linewidth}
         \begin{itemize}
             \item CAN IDs
             \item Data payload
         \end{itemize}
        \end{minipage}  & \begin{minipage}[t][1.1cm]{\linewidth}
         \begin{itemize}
             \item Efficient for \newline message level
         \end{itemize}
        \end{minipage}   & \begin{minipage}[t][1.1cm]{\linewidth}
         \begin{itemize}
             \item Sensitive to noise
             \item High computational cost
         \end{itemize}
        \end{minipage}  \\
         
         \cline{2-6} & \cite{Avatefipour2019AnLearning} &\begin{minipage}[t][1.85cm]{\linewidth}
         \begin{itemize}
             \item OSVM
         \end{itemize}
        \end{minipage}   & \begin{minipage}[t][1.1cm]{\linewidth}
         \begin{itemize}
             \item CAN IDs
             \item Data payload
         \end{itemize}
        \end{minipage}  & \begin{minipage}[t][1.1cm]{\linewidth}
         \begin{itemize}
             \item Novel \newline meta-heuristic optimization algorithm
         \end{itemize}
        \end{minipage}  & \begin{minipage}[t][1.1cm]{\linewidth}
         \begin{itemize}
             \item Train and test on a simulation dataset
         \end{itemize}
        \end{minipage}  \\ 
         
         
         \cline{2-6} & \cite{Wang2018AHTM} & \begin{minipage}[t][1.3cm]{\linewidth}
         \begin{itemize}
             \item HTM
         \end{itemize}
        \end{minipage}  & \begin{minipage}[t][1.2cm]{\linewidth}
         \begin{itemize}
             \item CAN IDs
             \item Data payload
         \end{itemize}
        \end{minipage}  & \begin{minipage}[t][1.1cm]{\linewidth}
         \begin{itemize}
             \item Novel distributed anomaly detection system
         \end{itemize}
        \end{minipage}  &\begin{minipage}[t][1.1cm]{\linewidth}
         \begin{itemize}
             \item High time \newline complexity
         \end{itemize}
        \end{minipage}   \\
         
         \cline{2-6} & \cite{Ashraf2020NovelSystems} &\begin{minipage}[t][1.1cm]{\linewidth}
         \begin{itemize}
             \item LSTM \newline autoencoder
         \end{itemize}
        \end{minipage}   & \begin{minipage}[t][1.5cm]{\linewidth}
         \begin{itemize}
             \item CAN IDs
             \item Data payload
         \end{itemize}
        \end{minipage}  & \begin{minipage}[t][1.1cm]{\linewidth}
         \begin{itemize}
             \item Solve both \newline internal and \newline external attacks
         \end{itemize}
        \end{minipage}  & \begin{minipage}[t][1.1cm]{\linewidth}
         \begin{itemize}
             \item No time complexity evaluation
         \end{itemize}
        \end{minipage}  \\ 
         \hline 
         \multirow{4}{*}{Hybrid} & \cite{Seo2018GIDS:Network} &\begin{minipage}[t][0.8cm]{\linewidth}
         \begin{itemize}
             \item GAN
         \end{itemize}
        \end{minipage} & \begin{minipage}[t][1.1cm]{\linewidth}
         \begin{itemize}\item CAN IDs\end{itemize}
        \end{minipage}  &\begin{minipage}[t][1.1cm]{\linewidth}
         \begin{itemize}
             \item Detect unknown \newline attacks
         \end{itemize}
        \end{minipage} &\begin{minipage}[t][1.1cm]{\linewidth}
         \begin{itemize}
             \item Low accuracy
         \end{itemize}
        \end{minipage}  \\ 
         \cline{2-6} & \cite{Song2021Self-SupervisedData} &\begin{minipage}[t][1.5cm]{\linewidth}
         \begin{itemize}
             \item LSTM-based GAN
             \item DCNN
         \end{itemize}
        \end{minipage}   &\begin{minipage}[t][1.1cm]{\linewidth}
         \begin{itemize}\item CAN IDs\end{itemize}
        \end{minipage}  &\begin{minipage}[t][1.1cm]{\linewidth}
         \begin{itemize}
             \item Novel approach for \newline data generation \newline and sampling
         \end{itemize}
        \end{minipage}   & \begin{minipage}[t][1.1cm]{\linewidth}
         \begin{itemize}
             \item Low accuracy
         \end{itemize}
        \end{minipage}  \\ 
          \cline{2-6} & \cite{Yang2021MTH-IDS:Vehicles} &\begin{minipage}[t][1.8cm]{\linewidth}
         \begin{itemize}
              \item Supervised tree-based models
              \item K-means
          \end{itemize}
        \end{minipage} &\begin{minipage}[t][1.1cm]{\linewidth}
         \begin{itemize}
             \item CAN IDs
             \item Data payload
         \end{itemize}
        \end{minipage} &\begin{minipage}[t][1.1cm]{\linewidth}
         \begin{itemize}
             \item Detect unknown \newline attacks
         \end{itemize}
        \end{minipage} &\begin{minipage}[t][1.1cm]{\linewidth}
         \begin{itemize}
             \item Complicated data \newline preprocessing
         \end{itemize}
        \end{minipage}   \\
         \hline
    \end{tabular}    
    
    \label{tab:litrev}
\end{table*}

\section{Methodology \label{sec:method}}

We propose our methodology to address the limitations of related works, such as low detection rate for unknown attacks in supervised models and high false positive rate in unsupervised models. The proposed deep learning-based IDS is developed from an adversarial autoencoder (AAE) architecture, which is a combination of autoencoder (AE) and generative adversarial networks (GAN). The AAE scheme is suitable for in-vehicle IDS for two reasons. First, the AAE-based model can handle the data scarcity problem. This is because it does not require a large amount of data, which consumes a long time to collect and label. Particularly, safety has the highest priority in the vehicle domain. Therefore, the data source for attack samples is limited. Second, the AAE-based model can detect unknown attacks. Because the security of the CAN-bus system is extremely weak, the system is vulnerable to various types of attacks, which are updated frequently by intelligent adversaries. In some cases, security researchers are unaware of a new type of attack, therefore it is not labeled. 

In this section, we first explain the fundamental knowledge about AE, GAN, and AAE. Then, the details of our proposed system are presented.

\subsection{Autoencoder}

\begin{figure}[h!]
\centering
\includegraphics[width=2.5in]{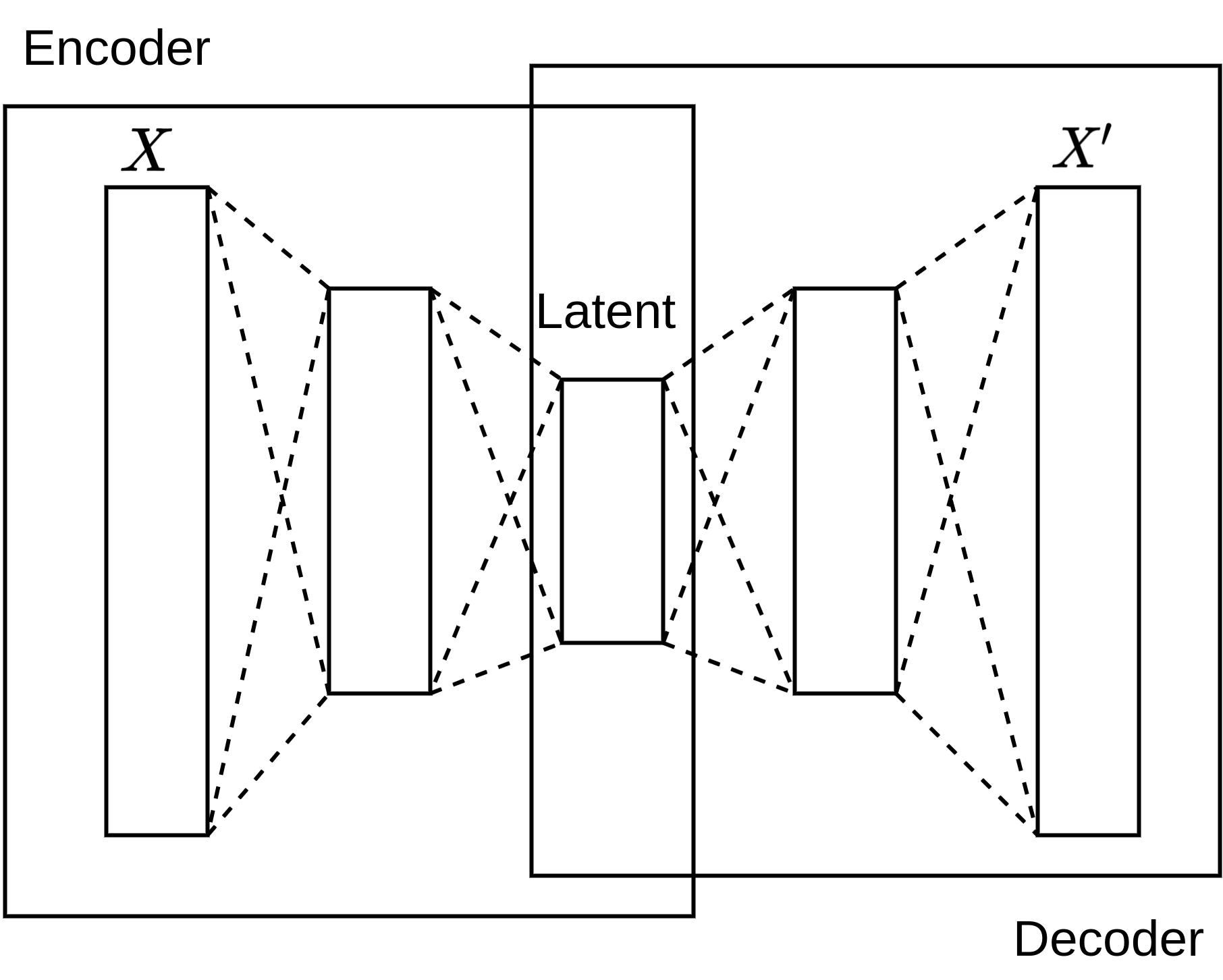}   
\caption{Autoencoder architecture.}
\label{fig:AE}
\end{figure}

An autoencoder (AE) \cite{Ian2016DeepLearning} is an unsupervised neural network that is trained to reconstruct its input. The network (\cref{fig:AE}) consists of two parts: an encoder maps the input to latent features and a decoder attempts to reconstruct the input with the encoder output. With the condition that the dimensionality of the latent space is assumed to be much lower than the dimensionality of the input, the autoencoder can learn useful properties of the data. Therefore, it is usually used for dimensionality reduction. The goal of the autoencoder is to minimize the reconstruction loss $L_R$, which can be defined as the squared error between the input $X$ and the reconstructed output $X'$ with $N$, the number of samples, as follows:

\begin{equation}\label{eq:reconstruction_loss}
L_R(\textbf{X}, \textbf{X}^\prime) = \frac{1}{N}||\textbf{X} - \textbf{X}^\prime||^2.    
\end{equation}

\subsection{Generative Adversarial Networks}

\begin{figure}[h!]
\centering
\includegraphics[width=3.5in]{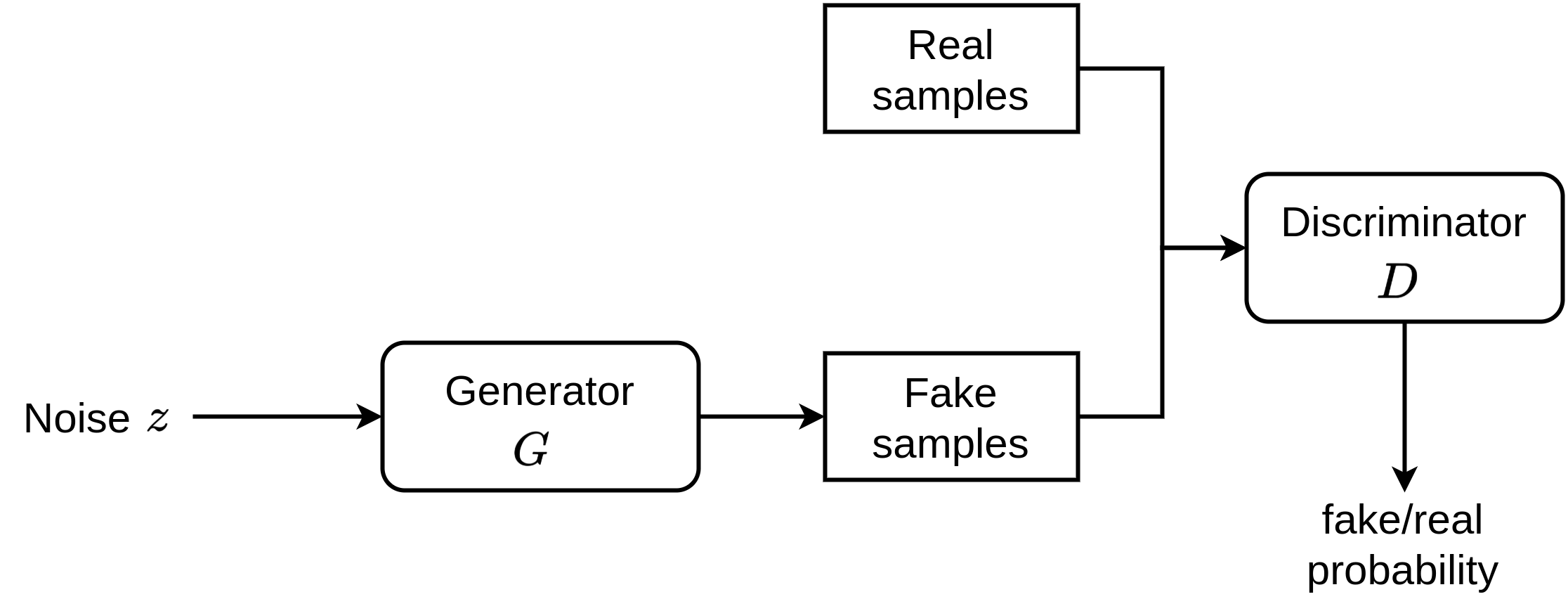}   
\caption{GAN architecture.}
\label{fig:GAN}
\end{figure}

Generative Adversarial Networks (GAN) \cite{GoodfellowGenerativeNets} is a deep learning-based model that uses a training set assumed to follow a distribution $p_{data}$. GAN learns to represent an estimate of that distribution and returns a probability distribution $p_{model}$ as the result. As illustrated in \cref{fig:GAN}, GAN is composed of two deep neural networks: generator ($G$) and discriminator ($D$). Although $G$ attempts to generate new data plausibly, $D$ is trained to distinguish real data derived from the training dataset and generated data from $G$.

To train $G$, we obtain the noise input {$\boldsymbol{z}$} following a predefined distribution. $G(\boldsymbol{z}; \boldsymbol{\theta_g})$ is new data generated from $G$ with parameter $\boldsymbol{\theta_g}$. Furthermore, $D(\boldsymbol{x}; \boldsymbol{\theta_d})$ represents the probability that $\boldsymbol{x}$ is real or fake, and $D(G(\textbf{z}))$ evaluates the authenticity of data generated from $G$. 

In addition, $D$ is trained to maximize the probability of assigning the correct label to both training examples and samples from $G$. The loss function of $D$ can be formulated as follows:
\begin{equation}
\begin{split}
    \mathop{\text{max}}_{D} V(D) &= \mathbb{E}_{\boldsymbol{x} \sim p_{data}(\boldsymbol{x})}[\text{log}(D(\boldsymbol{x}))] \\
    & + \mathbb{E}_{\boldsymbol{z} \sim p_{\boldsymbol{z}}(\boldsymbol{z})} [ \text{log}(1 - D(G(\boldsymbol{z}))].
\end{split}
\end{equation}
By contrast, $G$ wants to create new data that appear similar to the training data to decieve $D$. As a result, $G$ minimizes $log(1 - D(z))$. The loss function of $G$ is
\begin{equation}
\mathop{\text{min}}_{G} V(G) = \mathbb{E}_{\boldsymbol{z} \sim p_{\boldsymbol{z}}(\boldsymbol{z})} [1 - \text{log}(D(G(\boldsymbol{z}))].    
\end{equation}
To summarize, GAN loss can be written as
\begin{equation} \label{eq:gan_loss}
\begin{split}
  \mathop{\text{min}}_{G} \mathop{\text{max}}_{D} (V(D, G)) 
  & = \mathbb{E}_{\boldsymbol{x} \sim p_{data}(x)}[\text{log}(D(\boldsymbol{x}))] \\
  & + \mathbb{E}_{\boldsymbol{z} \sim p_z(\boldsymbol{z})} [1 - \text{log}(D(G(\boldsymbol{z}))].    
\end{split}
\end{equation}

\subsection{Adversarial Autoencoder}

\begin{figure}[h!]
\centering
\includegraphics[width=3in]{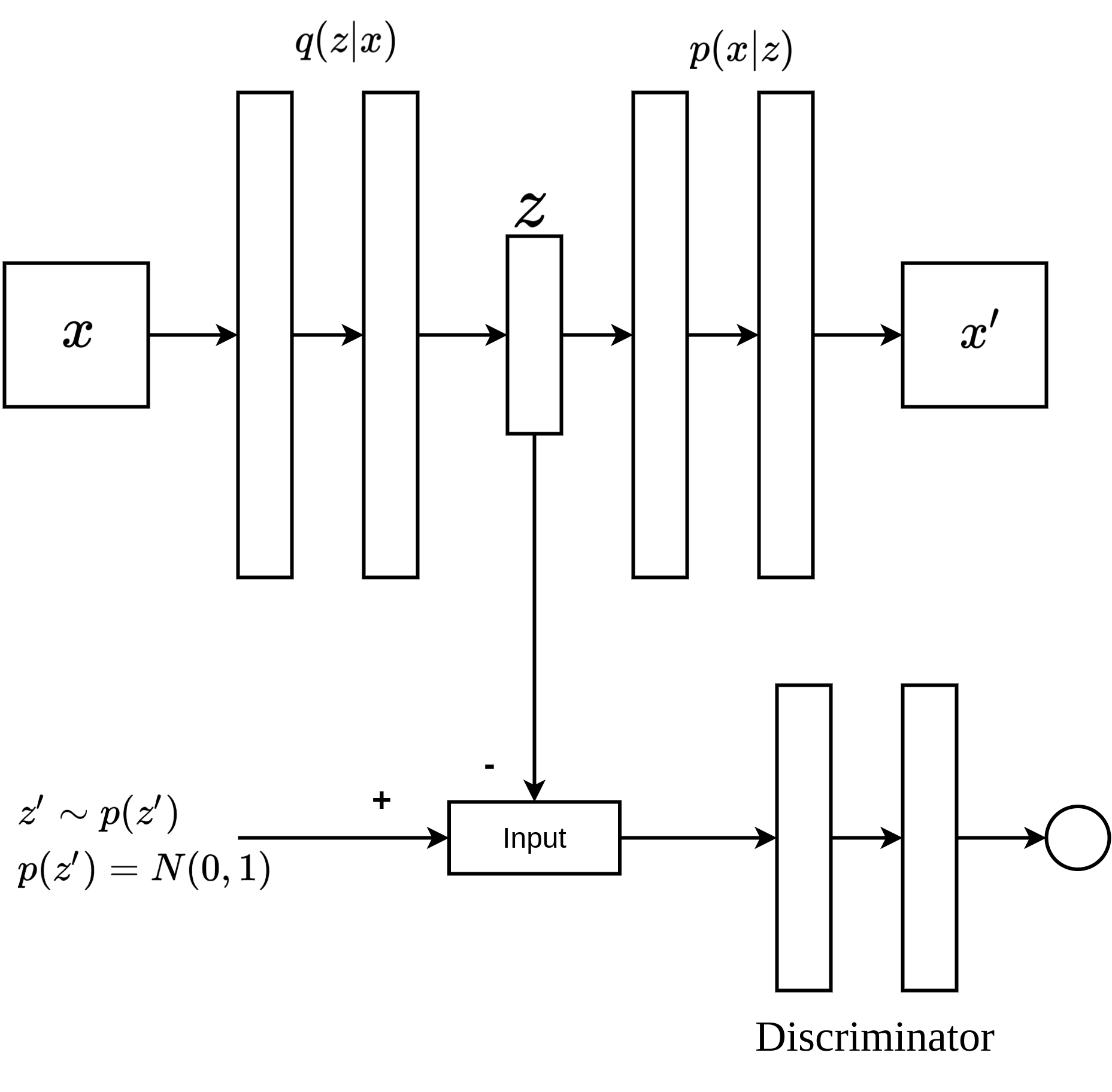}   
\caption{AAE architecture.}
\label{fig:AAE}
\end{figure}

AAE \cite{Makhzani2015AdversarialAutoencoders} is a technique to regularize a vanilla autoencoder by applying adversarial training. The goal is to force the latent space to follow an arbitrary distribution. The overview of AAE architecture is shown in \cref{fig:AAE}.
For example, we assume $\boldsymbol{x}$ is the input and $\boldsymbol{z}$ is the latent variable of the AE. Furthermore, assuming $p(\boldsymbol{z})$ is the prior distribution that we want to impose on the latent variable, $q(\boldsymbol{z}|\boldsymbol{x})$ is an encoding distribution, and $p(\boldsymbol{x}|\boldsymbol{z})$ is a decoding distribution. The encoding function of the AE, $q(\boldsymbol{z}|\boldsymbol{x})$, defines an aggregated posterior distribution of $q(\boldsymbol{z})$ on the latent variable of the AE as follows:

\begin{equation} \label{eq:aae_loss}
\begin{split}
  q(\boldsymbol{z}) = \int_x q(\boldsymbol{z}|\boldsymbol{x})p(\boldsymbol{x}|\boldsymbol{z})dx.    
\end{split}
\end{equation}

Thus, the AAE can match the prior distribution $p(\boldsymbol{z})$ to the aggregated posterior $q(\boldsymbol{z})$ of latent variable $\boldsymbol{z}$. In other words, the latent variable $\boldsymbol{z}$ follows the prior distribution, and the encoder of AE plays as the generator of GAN, which generates the latent variable $\boldsymbol{z}$. The discriminator scores the similarity between $\boldsymbol{z}$ and samples $\boldsymbol{\boldsymbol{z^\prime}}$ drawn from prior distribution of $p(\boldsymbol{z})$. AAE is trained with stochastic gradient descent (SGD) in two phases as follows:

\begin{enumerate}
    \item Reconstruction phase: only the autoencoder part is trained with reconstruction loss in this phase. First, the input is fed into the encoder to obtain the latent features. The decoder decodes those latent features to retrieve the input. This phase will optimize the reconstruction loss, $L_R$ defined in \eqref{eq:reconstruction_loss}. 
    \item Regularization phase: the generator (encoder) and discriminator are trained together. First, the discriminator learns how to classify the encoder output and random input from the prior distribution. The random input is labeled as 1, and the encoder output is labeled as 0. Then, we fix the discriminator and train the generator to produce output following the prior distribution. To achieve that, the generator is trained to minimize the cross-entropy between target values, which are all set to be 1, and the scores that the discriminator provides to the encoder outputs. This phase attempts to minimize the GAN loss defined in \eqref{eq:gan_loss} by replacing $\boldsymbol{x}$ to $\boldsymbol{\boldsymbol{z^\prime}}$.
\end{enumerate}

\subsection{Proposed model}

\begin{figure}[h!]
\centering
\includegraphics[width=3in]{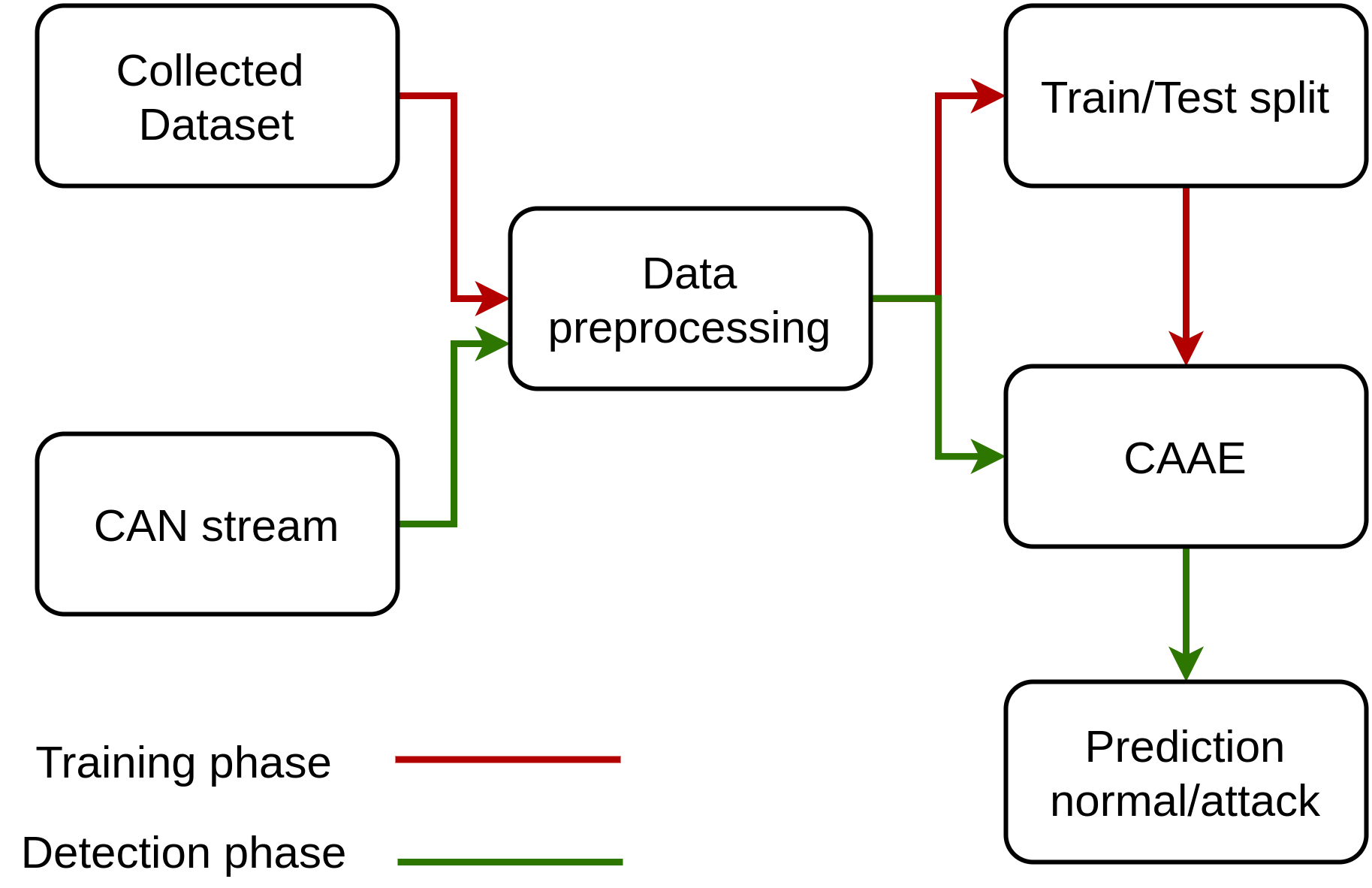}   
\caption{Workflow overview.}
\label{fig:workflow}
\end{figure}

\subsubsection{Workflow overview}

The workflow (\cref{fig:workflow}) consists of 2 phases: training and testing. We propose a deep learning model named the convolutional adversarial autoencoder (CAAE). As suggested by \cite{Song2020In-vehicleNetwork}, we utilize only the CAN ID in messages to capture its sequential pattern. After data is collected, we construct a frame with a size of $29 \times 29$ by stacking 29 consecutive CAN IDs in a 29-bit representation. Then, the frame is fed into the CAAE model.  

In the training phase, we label the frame as abnormal if there is at least one injected message. However, we do not need to label all of them because our model only needs a small number of labeled data, which helps us save time on labeling frames as well as messages. The training process will be completed offline, whereas the detection occurs online to serve in real-time. 

\begin{figure*}[H!]
\centering
\includegraphics[width=5in]{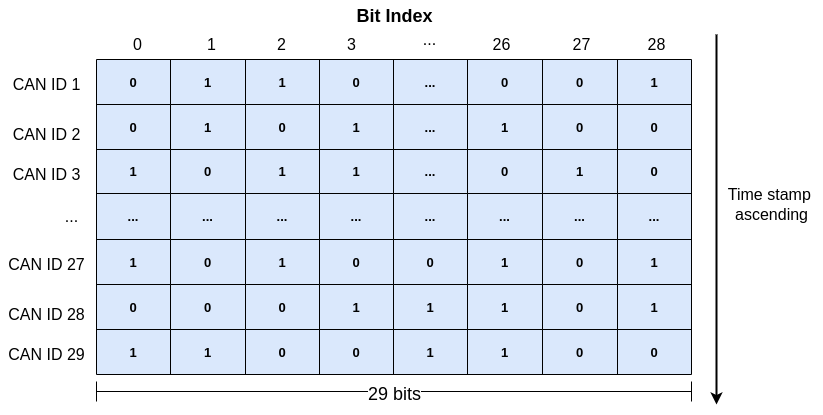}   
\caption{CAN ID frame - input to CAAE model}
\label{fig:frame}
\end{figure*}

\subsubsection{Data preprocessing}

Based on the working principles of the CAN bus system, the proposed method utilizes the CAN IDs as features because there is a pattern in the CAN IDs sequence. Whenever an injected message exists, the pattern will be broken. Therefore, using the CAN IDs sequence, the model can capture the normal and abnormal patterns and classify them correctly. To enable the model to adapt to any version of CAN messages and increase efficiency, the CAN IDs are represented in 29 bits, which means that each CAN ID is illustrated as follows:

\begin{equation}
    \text{ID} = b_i (\text{for } i = 0, ..., 28),
\end{equation}
where $b_i$ is the bit at $i^{th}$ position. The \cref{fig:frame} illustrates a frame that is constructed by stacking 29 continuous CAN IDs together. We chose 29 because a square matrix is easier for CNN to process the input data. The frames are the inputs fed into the CAAE model, which will be described in detail in the next section.

\subsubsection{Convolutional Adversarial Autoencoder (CAAE)}

\begin{figure}[h!]
\centering
\includegraphics[width=3.3in]{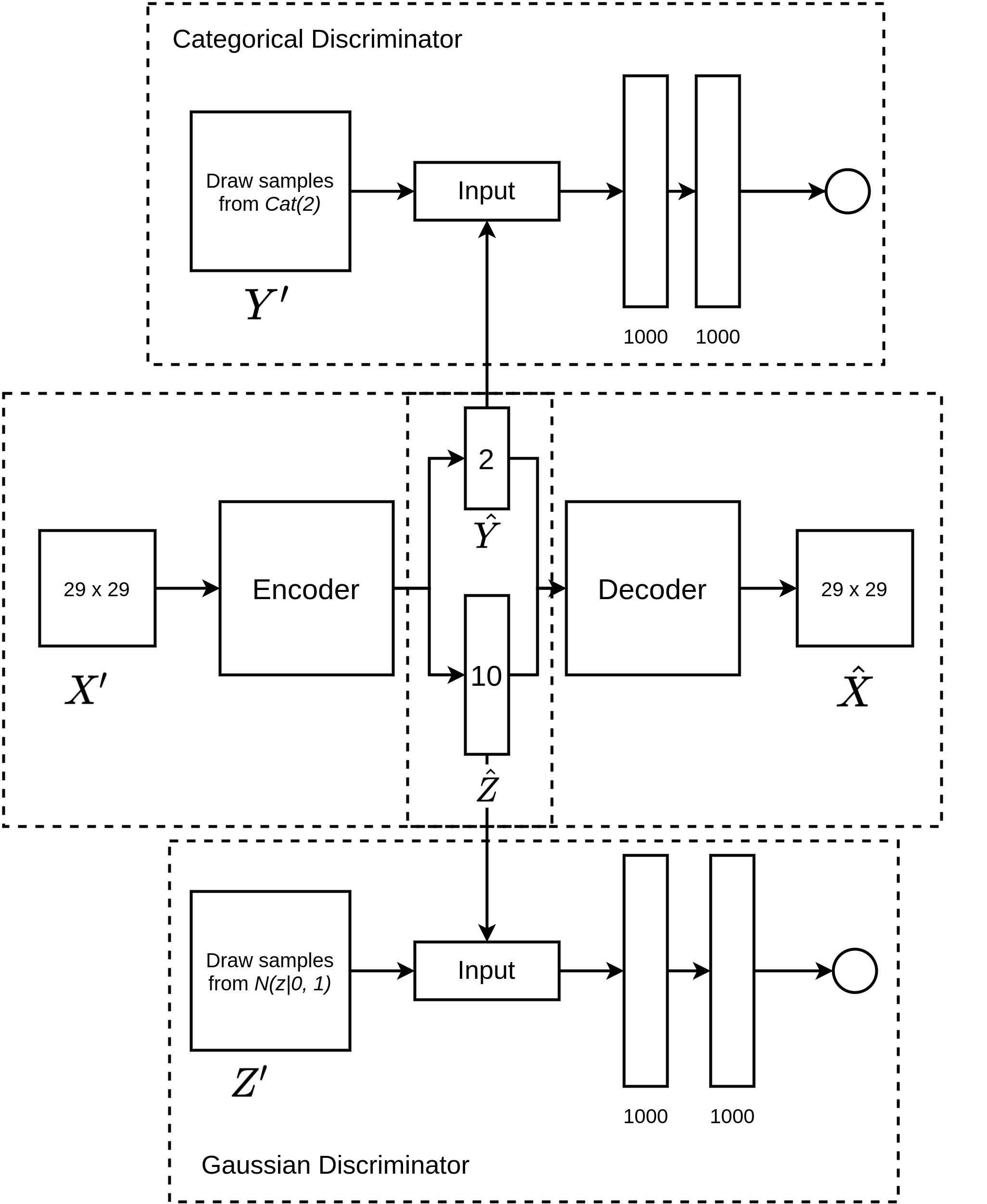}   
\caption{CAAE architecture.}
\label{fig:CAAE}
\end{figure}

 To reduce the amount of labeled data, we train the CAAE in semi-supervised learning, which combines a small amount of labeled data with a large amount of unlabeled data during training. The architecture of our model is presented in \cref{fig:CAAE}. The input consists of $n$ labeled samples denoted as $\{\boldsymbol{X_l}, \boldsymbol{Y_l}\}$ and $m$ unlabeled samples denoted as $\{\boldsymbol{X_{ul}}\}$, where $n \ll m$.  
The encoder generates two latent variables: $\boldsymbol{\widehat{Y}} \in \mathbb{R}^2$ is for class information (normal and abnormal) and $\boldsymbol{\widehat{Z}} \in \mathbb{R}^{10}$ is for other features. Therefore, the proposed model needs two discriminators: $D_{cat}$ forces $\boldsymbol{\widehat{Y}}$ to follow the categorical distribution $Cat(2)$, whereas $D_{gaus}$ forces $\boldsymbol{\widehat{Z}}$ to follow the Gaussian distribution with the mean of zero and identity covariance. Because the representations of normal and abnormal patterns can be very complicated and cannot be described by only the class information $\boldsymbol{\widehat{Y}}$, we use $\boldsymbol{\widehat{Z}}$ to preserve other specific features, and then samples are mapped accurately in the latent space.

The training process for CAAE is similar to that for AAE, except 
that a supervised phase is added. First, we train the autoencoder part by feeding a batch of $\boldsymbol{\{X_{ul}\}}$. The next step is training two discriminators corresponding to each type of latent feature and the encoder in an adversarial way. Each discriminator comprises two layers, with 1000 neurons for each layer. The discriminator output is the probability generated by the sigmoid activation function. To make the model more stable and converge faster, we use the Wasserstein GAN (WGAN) loss with gradient penalty (GP) \cite{GulrajaniImprovedAlgorithms}. 
Rather than using Jensen–Shannon divergence, WGAN uses Wasserstein distance to calculate the difference between real and fake distributions. We consider the categorical discriminator as an example. The categorical discriminator $D_{cat}$ attempts to minimize the loss defined as follows: 
\begin{equation} \label{eq:wgan}
    L_{WGAN} = \mathbb{E}[D_{cat}(\boldsymbol{\widehat{Y}})] - \mathbb{E}[D_{cat}(\boldsymbol{Y^\prime})],
\end{equation}
where $\boldsymbol{Y^\prime}$ is the samples drawn from the categorical distribution $Cat(2)$.
A gradient penalty, which is added to the loss to ensure the 1-Lipschitz constraint in GAN, is defined as follows:
\begin{equation} \label{eq:gp}
    \text{GP}_{cat} = \mathbb{E}[(\| \nabla_{\boldsymbol{\tilde{Y}}} D(\tilde{Y}) \| - 1)^2],
\end{equation}
where $\tilde{Y} = \epsilon \widehat{Y} + (1 - \epsilon) \boldsymbol{Y^\prime}$, $\epsilon$ is a random variable, and $\epsilon \sim \mathcal{U}[0,1]$. The final loss is the sum of (\ref{eq:wgan}) and (\ref{eq:gp})
\begin{equation}
    L_{cat} = L_{WGAN} + \lambda \text{GP}_{cat},
\end{equation}
where $\lambda$ is a penalty factor and is usually set to 10 \cite{GulrajaniImprovedAlgorithms}. These formulas are similar when applied for Gaussian discriminator $D_{gaus}$.
Following this step, the model learns to extract useful features from unlabeled data. Finally, we train the encoder with labeled samples $\{\boldsymbol{X_l}, \boldsymbol{Y_l}\}$ by minimizing the cross-entropy loss. The aforementioned steps are summarized in \cref{alg:caae}, where $\boldsymbol{\theta}$ indicates the parameters of the model. Although the training process appears complicated, only the encoder's weights are saved and utilized for online detection. Consequently, our model is lightweight and efficient for the in-vehicle IDS.   

\begin{figure*}[h!]
\centering
\includegraphics{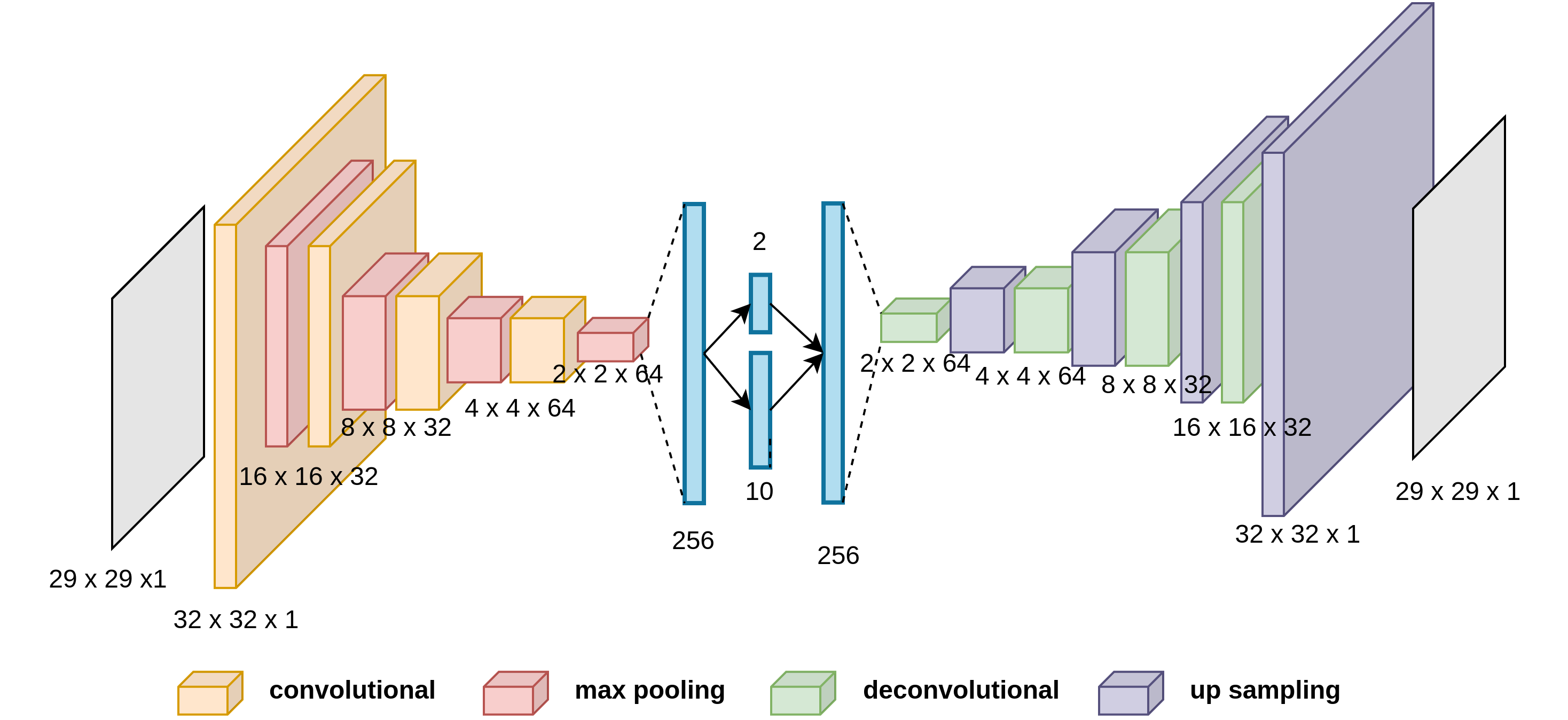}   
\caption{Detail of convolutional layers in Autoencoder.}
\label{fig:CAE}
\end{figure*}

\begin{algorithm}
\caption{CAAE training process}\label{alg:caae}
\textbf{Input:} 

\begin{itemize}
    \item $\boldsymbol{\{X_{ul}\}}$: Unlabaled dataset. 
    \item ${\{\boldsymbol{X_l, Y_l}\}}$: Labeled dataset.  
    \item $n\_epochs$: Number of epochs.  
    \item $batch\_size$: Batch size. 
\end{itemize}

\textbf{Output:} 

\begin{itemize}
    \item $\boldsymbol{\theta}_{En}$: Encoder's weights. 
\end{itemize}

\begin{algorithmic}[1]
    \FOR{$epoch$:= 1 \TO $n\_epochs$}
        \STATE Sample minibatch for $\{\boldsymbol{X_l}, \boldsymbol{Y_l}\}$, $\boldsymbol{\{X_{ul}\}}$.
        \STATE Draw samples for $\boldsymbol{Y^\prime} \sim Cat(2)$, $\boldsymbol{Z^\prime} \sim N(0, I_{10})$.
        \STATE \textbf{Reconstruction phase}
        \STATE  Update $\boldsymbol{\theta}_{En}$ and $\boldsymbol{\theta}_{De}$ by minimizing
        
        \(L_R = \frac{1}{batch\_size}\|\boldsymbol{X}_{ul} - \boldsymbol{\widehat{X}}\|^2\).
        
        \STATE \textbf{Regularization phase}
        \STATE  Update $\boldsymbol{\theta}_{D_{cat}}$ by minimizing 
        
        \(L_{cat} = \mathbb{E}[D_{cat}(\boldsymbol{\widehat{Y}})] - \mathbb{E}[D_{cat}(\boldsymbol{Y^\prime})] + \lambda \text{GP}_{cat}\).
        
        \STATE  Update $\boldsymbol{\theta_{D_{gaus}}}$ by minimizing 
        
        \(L_{gaus} = \mathbb{E}[D_{gaus}(\boldsymbol{\widehat{Z}})] - \mathbb{E}[D_{gaus}(\boldsymbol{Z^\prime})] + \lambda \text{GP}_{gaus}\).
        
        \STATE Update $\boldsymbol{\theta}_{En}$ by minimizing 
        
        \(L_{gen} = -\mathbb{E}[D_{cat}(\boldsymbol{\widehat{Y}})] - \mathbb{E}[D_{gaus}(\boldsymbol{\widehat{Z}})]\).
        
        \STATE \textbf{Supervised learning phase}
        
        \STATE Using $\{\boldsymbol{X_l}, \boldsymbol{Y_l}\}$ to update $\boldsymbol{\theta}_{En}$ by minimizing
        
        \(L_{sup} = -\mathbb{E}[\text{log}(\boldsymbol{\widehat{Y}}) * \boldsymbol{Y_l}]\).
    \ENDFOR
\end{algorithmic}
\end{algorithm}

\begin{table}[t!]
    \centering
    \caption{Hyperparameter value of the CAAE model}
    \begin{tabular}{c|c}
    \hline
        Hyperparameter &  Value \\
        \hline Batch size   & 64 \\ 
        Supervised learning rate & $10^{-4}$\\
        Reconstruction learning rate & $10^{-4}$ \\
        Regularization learning rate &  $10^{-4}$ \\
        Decay for learning rate update & 0.1 \\
        Latent space dimension & 2 + 10 \\
        \hline
    \end{tabular}
    \label{tab:hyperparameter}
\end{table}

Convolution, which is well-established for spatial and sequential patterns, is extremely suitable for the CAN IDs data. Therefore, we added 2D convolutional layers to our autoencoder structure. Because we trained the model in a semi-supervised manner, our convolution neural network is very simple. Thus, the proposed model still adapts to the real-time requirement for in-vehicle IDS. The $29 \times 29$ frame input is transformed into a $32 \times 32$ input by padding. This is because an even-numbered size is more suitable for the convolutional autoencoder. For the encoder, we applied convolution with a kernel size of $3 \times 3$ and max-pooling layers. Then the result is flattened and applied to two fully connected networks to generate two types of latent features. Then, the encoder outputs are concatenated before being fed into the decoder. By contrast, the decoder includes deconvolution layers with the same kernel size as the encoder and upsampling layers. The final result is cropped to regain the $29 \times 29$ frame. To prevent overfitting during training, we added a dropout layer \cite{Srivastava2014Dropout:Overfitting} with a rate of 0.15 before the fully connected layers. In addition, we used ReLU \cite{Nair2010RectifiedMachines} for activation functions and the ADAM optimizer \cite{Kingma2014Adam:Optimization} for backpropagation. We also used the learning rate decay technique, which decreases the learning rate by 10 after the $50^{th}$ epoch. The details of the architecture of the convolutional autoencoder and hyperparameters for training the CAAE model are shown in \cref{fig:CAE} and \cref{tab:hyperparameter} respectively.

\section{Experimental results \label{sec:results}}

\subsection{Datasets}

\begin{table}[t!]
    \centering
    \caption{Car hacking dataset overview}
    \begin{tabular}{c|c|c}
         \hline Attack type & Normal messages & Injected messages  \\
         \hline DoS Attack & 3,078,250 (84\%) & 587,521 (16\%) \\ 
          Fuzzy Attack & 3,347,013 (87\%) & 491,847 (13\%) \\ 
          Gear Spoofing & 2,766,522 (82\%) & 597,252 (18\%) \\ 
          RPM Spoofing & 2,290,185 (78\%) & 654,897 (22\%) \\
          \hline
    \end{tabular}
    \label{tab:datasets}
\end{table}

\begin{table}[t!]
    \centering
    \caption{Preprocessed datasets overview.}
    \begin{tabular}{c|c}
         \hline Attack type & Frames  \\
         \hline Normal & 352,767 (62\%) \\
          DoS Attack & 37,451 (7\%) \\ 
          Fuzzy Attack & 44,486 (8\%) \\ 
          Gear Spoofing & 65,283 (11\%) \\ 
          RPM Spoofing & 71,372 (12\%) \\
          \hline
    \end{tabular}
    \label{tab:dataframes}
\end{table}

We used the car hacking datasets \cite{Song2020In-vehicleNetwork} produced by the Hacking and Countermeasure Research Lab (HCRL) of Korea University. The dataset was constructed by logging CAN traffic via the OBD-II port of a real vehicle while malfunctioning messages are injected. There are four types of attacks: DoS, fuzzy, spoofing RPM, and spoofing gear information, which are saved in the different comma-separated value files. Table \ref{tab:datasets} shows the details of this dataset. 

Each message includes timestamp, CAN ID in HEX, the number of data bytes, 8-byte data, and a flag with two values, which are T for an injected message and R for a normal message. We extracted CAN IDs and transformed them from hexadecimal to a 29-bit representation. Then, the data frame was constructed by stacking 29 sequential samples together as shown in \cref{fig:frame}. In our model, there are 2 classes: normal (0) and abnormal (1). The frame was labeled as abnormal if there was at least one injected message. Table \ref{tab:dataframes} shows the information about data frames after preprocessing. 

\subsection{Experiment setup}

With normal frames, we divided the data frames into the training set, validation set, and test set with the percentage of 70\%, 15\%, and 15\%, respectively. The validation set was used for checking the overfitting and tuning hyperparameters during training. The test set was kept hidden until the training progress is complete. To demonstrate that our model can detect intrusion activities with a small number of attack samples, we adjusted the number of attack samples. We used 10\%, 30\%, 50\%, and 70\% of total attack data for training. Moreover, the labeled data comprised only 10\% of the total training data. Table \ref{tab:dataset_structure} displays the details of the training dataset structure.

The computer configuration for the experiment included a 64-bit Intel (R) Core(TM) i7-7700 CPU @ 3.6 GHz and a Geforce GTX 1060 6GB GPU. GPU was used only for accelerating the training process. We trained and tested the proposed model with Tensorflow version 1.15 and Python 3.6.

\subsection{Evaluation metrics}

Because the test set can be imbalanced between normal and abnormal data, we used precision, recall, and F1 score to evaluate the performance of our model. In addition, the error rate (ER) is very important in the in-vehicle IDS. For example, if the IDS results in a high false negative rate, it can be dangerous for the driver; or if the IDS produces a high false positive rate, it can affect the user experience. Our goal is to achieve a high F1 score and a low ER. These metrics can be calculated by using true positive (TP), true negative (TN), false positive (FP), and false negative (FN):

\begin{equation}
    \text{Error rate (ER)} = \frac{FP + FN}{TP + TN + FP + FN}
\end{equation}

\begin{equation}
    \text{Recall (Rec)} =  \frac{TP}{TP + FN}
\end{equation}

\begin{equation} 
    \text{Precision (Pre)} = \frac{TP}{TP + FP}
\end{equation}

\begin{equation}
    \text{F1-score (F1)}  = 2\times \frac{Prec \times Rec}{Prec + Recall}.
\end{equation}

\begin{table}[t]
    \caption{Training settings}
    \begin{threeparttable}
    \begin{tabular}{c|c|c|c|c|c}
        \hline
        \multirow{2}{*}{Train ratio\tnote{*}} & \multicolumn{5}{c}{Labeled} \\
        \cline{2-6} & DoS & Fuzzy & Gear & RPM & Total \\
        \hline  0.1 & 400 & 450 & 650 & 700 & 2.20k \\ 
         0.3 & 1.20k & 1.35k & 1.95k & 2.10k & 6.60k \\ 
         0.5 & 2.00k & 2.25k & 3.25k & 3.50k & 11k \\ 
         0.7 & 2.80k & 3.15k & 4.55k & 4.90k & 15.4k \\ 
        \hline
    \end{tabular}
    \begin{tablenotes}
        \item[*] The train ratio indicates the number of training samples over the total samples in the dataset.
    \end{tablenotes}
    \end{threeparttable}
    \label{tab:dataset_structure}
\end{table}

\begin{table}[t]
    \centering
    \caption{Detection results for different amounts of attack data for known attacks}
    \begin{threeparttable}
    \begin{tabular}{c|c|c|c|c}
         \hline Train ratio\tnote{*} & ER & Rec & Prec & F1 \\
         \hline 0.1 & 3.2\% & 0.9620 & 0.9999 & 0.9806 \\ 
          0.3 & 0.9\% & 0.9893 & 0.9997 & 0.9945 \\ 
          0.5 & 1.0\% & 0.9821 & 0.9985 & 0.9902 \\ 
          0.7 & 0.4\% & 0.9899 & 0.9996 & 0.9947 \\
         \hline 
    \end{tabular}
    \begin{tablenotes}
        \item[*] The train ratio indicates the number of training samples over the total samples in the dataset.
    \end{tablenotes}
    \end{threeparttable}
    \label{tab:detection_result}
\end{table}

\begin{table}[t]
    \centering
    \caption{Detection results for different amounts of labeled data for known attacks}
    \begin{threeparttable}
    \begin{tabular}{c|c|c|c|c}
         \hline Label ratio\tnote{*} & ER & Rec & Prec & F1 \\
         \hline 0.1 & 0.4\% & 0.9899 & 0.9996 & 0.9947 \\ 
          0.2 & 0.2\% & 0.9942 & 0.9998 & 0.9970 \\
          0.3 & 0.2\% & 0.9958 & 0.9998 & 0.9978 \\ 
          0.4 & 0.1\% & 0.9972 & 0.9997 & 0.9984 \\
         \hline 
    \end{tabular} 
    \begin{tablenotes}
        \item[*] The label ratio indicates the number of labeled samples over the number of training samples.
    \end{tablenotes}
    \end{threeparttable}
    \label{tab:labeled_detection_results}
\end{table}

\subsection{Detection for known attacks}

In this section, the results of detection for known attacks are investigated. First, we attempt with a different amount of attack data where only $10\%$ of the data is labeled. Next, we also test the ability to detect the model when the number of labeled training data is increased. Finally, we compare our scheme with other supervised models.

\subsubsection{Results for known attacks}

Table \ref{tab:detection_result} shows the result of detection phase with different settings. The training ratio is defined as the ratio of the number of training samples to the total number of samples in the dataset. Therefore, the greater the number of samples, the better the detection results. For example, the ER decreases by 2.8\% if we change the training ratio from 10\% to 70\%. This is because 10\% of the total data is a very small number, which is not sufficient for the model to learn. The model achieves the best result of 0.4\% of ER when we use 70\% of the total amount of attack data for training. This result is very impressive because we only used 10\% labeled data. Moreover, the results can be improved if we increase the labeled data, which will be shown in the following experiment.



To evaluate the impact of the amount of labeled data, we used $70\%$ of total attack data for training and adjusted the labeled ratio with $10\%$, $20\%$, $30\%$, and $40\%$. We defined the labeled ratio as the fraction between the number of labeled training samples over the total training samples. As shown in Table \ref{tab:labeled_detection_results}, the more the number of labeled data, the lower ER and the higher F1 score. Moreover, the recall increased significantly from 0.9899 to 0.9972 if we increased the labeled ratio from 0.1 to 0.4, whereas the precision was considerably stable. The results also indicated that the false negative was reduced when more labeled attack data were fed. The model achieved $0.1\%$ and 0.9984 in terms of ER and F1 score with $70\%$ of total data in which $40\%$ of them were labeled. The result is very competitive to other supervised methods, which will be discussed in the next section.

\subsubsection{Comparison with other supervised methods}

Table \ref{tab:cmp_ml} presents the comparison between our model and other machine learning algorithms. The proposed model is trained with 70\% of total attack data, of which 40\% of the data were labeled. In addition, the other supervised models were trained with 100\% labeled data. The data processing for all these models is the same. We chose these models for diversification purposes: SVM for a kernel-based model, DT for a tree-based model, ANN for a neural network model, and DCNN for a deep learning model.

First, compared to the simple models, such as SVM, DT, and ANN, the results show that our model achieved the lowest ER and the highest F1 score. Most traditional machine learning models have low recall because they usually suffer from imbalanced classes. DT model performs the worst, with an ER of 1.77\% and an F1 score of approximately 0.98.
In addition, the ERs of the SVM and ANN models are 0.21\% and 0.15\%, respectively. Although the results of the two models are slightly worse than ours, it is noticeable that they are trained with 100\% labeled data, whereas our model used only 40\% of them. The next step is the comparison between the proposed model and the DCNN model, which is the state-of-the-art for the in-vehicle IDS. 

Interestingly, there is no significant difference between the CAAE model and the DCNN model, with 0.03\% of ER and 0.0007 of F1 score. It should be noted that our model used only 40\% labeled data, whereas DCNN used 100\% labeled data. The result suggests that the CAAE model can save time and effort for collecting and labeling data considerably, particularly when a new attack occurs. Moreover, the proposed model can detect unknown attacks, whereas DCNN and other traditional machine learning models cannot, because these models are trained in supervised learning, which can only classify patterns existing in training samples.

\begin{table}[t]
    \centering
    \caption{Comparison to other supervised methods for known attacks}
    \begin{threeparttable}
    \begin{tabular}{M{2cm}|c|c|c|c}
         \hline Model & ER & Rec & Prec & F1 \\
         \hline SVM & 0.21\% & 0.9947 & 0.9998 & 0.9972 \\ 
          Decision Tree & 1.77\% & 0.9799 & 0.9740 & 0.9770 \\ 
          ANN & 0.15\% & 0.9962 & 0.9999 & 0.9980 \\
          DCNN \cite{Song2020In-vehicleNetwork} & 0.07\% & 0.9984 & 0.9998 & 0.9991 \\ 
          Ours \tnote{*} & 0.1\% & 0.9972 & 0.9997 & 0.9984 \\
         \hline Compared to the best & 0.03\% & -0.0012 & -0.0010 & -0.0007 \\
         \hline 
    \end{tabular}
    \begin{tablenotes}
    \item[*] Note: All the supervised models were trained with 100\% labeled samples, whereas our model used only 40\% of them.    
    \end{tablenotes}
    \end{threeparttable}
    \label{tab:cmp_ml}
\end{table}

\begin{table*}[ht]
    \centering
    \caption{Detection results for unknown attacks}
    \begin{tabular}{c|c|c|c|c|c|c}
        \hline
         \multirow{2}{*}{Unknown attack} & \multicolumn{3}{c|}{Unknown Results} & \multicolumn{3}{c}{Known Results}\\ 
         \cline{2-7} & Rec & Prec & F1 & Rec & Prec & F1 \\
         \hline \centering DoS &  0.9823 & 0.9992 & 0.9907 & 0.9865 & 0.9999 & 0.9931 \\ 
          \centering Fuzzy &  0.8426 & 0.9999 & 0.9145 & 0.9953 & 1.0 & 0.9976 \\ 
          \centering Gear Spoofing &  0.9978 & 0.9977 & 0.9977 & 0.9674 & 0.9992 & 0.9831 \\
          \centering RPM Spoofing &  0.9955 & 0.9984 & 0.9970 & 0.9789 & 0.9994 & 0.9890  \\
         \hline
    \end{tabular}
    \label{tab:unknown_result}
\end{table*}
\begin{table*}[ht]
    \centering
    \caption{Comparison for unknown attacks}
    \begin{tabular}{c|c|c|c|c}
         \hline Unknown attack & Model & Rec & Prec & F1  \\
         \hline 
         \multirow{3}{*}{DoS} & Deep Autoencoder \cite{Ca2010StackedCriterion}& \textbf{0.9988} & 0.9127 & 0.9538 \\ 
          & Self-supervised learning \cite{Song2021Self-SupervisedData}& 0.9916 & 0.9751 & 0.9833 \\ 
          & Ours &  0.9823  & \textbf{0.9992} & \textbf{0.9907} \\
         \hline 
         \multirow{3}{*}{Fuzzy} & Deep Autoencoder \cite{Ca2010StackedCriterion}& \textbf{0.9626} & 0.9005 & \textbf{0.9305} \\ 
          & Self-supervised learning \cite{Song2021Self-SupervisedData}& 0.8345 & 0.9445 & 0.8861 \\ 
          & Ours & 0.8426 & \textbf{0.9999} &  0.9145 \\
          \hline 
         \multirow{3}{*}{Gear Spoofing} & Deep Autoencoder \cite{Ca2010StackedCriterion}& 0.8180 & 0.9463 & 0.8775 \\ 
          & Self-supervised learning \cite{Song2021Self-SupervisedData}& 0.8803 & 0.9768 & 0.9261 \\ 
          & Ours &  \textbf{0.9978}  & \textbf{0.9977}  & \textbf{0.9977} \\
         \hline 
         \multirow{3}{*}{RPM Spoofing} & Deep Autoencoder \cite{Ca2010StackedCriterion}& 0.9573 & 0.9573 & 0.9573 \\ 
          & Self-supervised learning \cite{Song2021Self-SupervisedData}& \textbf{0.9997} & 0.9720 & 0.9850 \\ 
          & Ours &  0.9955 & \textbf{0.9984} & \textbf{0.9970} \\
         \hline 
    \end{tabular}
    \label{tab:cmp_unknown}
\end{table*}
\begin{table}[ht]
    \centering
    \caption{Model complexity comparison}
    \begin{tabular}{M{1cm}|M{2cm}|M{2cm}|M{2cm}}
    \hline \multirow{2}{*}{Model} & \#Parameters & \multicolumn{2}{c}{Inference time (ms)} \\
    \cline{3-4}& (million) & GPU & CPU \\ 
    \hline DCNN & 9.80 & 5.00 & 6.70  \\ 
                Ours & 2.15 & 0.63 & 0.69 \\ 
         \hline 
    \end{tabular}
    \label{tab:cmp_dcnn_time}
\end{table}
\subsection{Detection for unknown attacks}
We define an unknown attack as an attack that can stealthily occur in training data but is not labeled. Our model can detect this kind of attack. To test it, we considered each kind of attack as an unknown attack by eliminating its labeled data from the training data. For example, if the unknown attack is DoS, we trained the model with labeled data including normal, fuzzy, gear, and RPM attacks. In this case, we produced unknown results by using the test sets of normal and DoS attack; and known results by using the test sets of normal and the other types. It is similar to three other kinds of attacks. In this experiment, we used only 30\% of total data, of which 10\% of data were labeled.

Table \ref{tab:unknown_result} presents the results of unknown attack testing. Gear and RPM achieved the highest F1 scores of more than 0.99. Consequently, there is a possibility that gear and RPM all belong to the spoofing attacks. When one of them is removed, the model still can detect the other. By contrast, the recalls of DoS and fuzzy are 0.98 and 0.84, respectively. The results can be acceptable because the labeled data were not used. Furthermore, the results are evidence that the proposed model can detect unknown attacks. However, it can be observed that there is a trade-off between known and unknown attacks. There is only an F1 score of DoS attack achieve higher than 0.99 for both unknown and known tests. 

We compared our results with those of two other methods which can detect unknown attacks. The first model is a deep autoencoder (DAE) \cite{Ca2010StackedCriterion} trained with only normal data. The model detects attacks by checking whether the reconstruction loss of the new sample is higher than a predefined threshold. The second model is the self-supervised learning method presented in \cite{Song2021Self-SupervisedData}. The model is trained with normal and generated data, which were labeled as attack samples. An additional RPM data is used as hint data to improve the final results. 

As shown in \cref{tab:cmp_unknown}, the proposed method achieves the highest F1 score in different kinds of attacks, except the fuzzy attack. In addition, the DAE model achieved an impressive F1 score of 0.9305 for fuzzy attack. However, the other results of this model are not good, with the worst F1 score at 0.8775 for gear spoofing attack. In addition, the DAE model achieved the recall higher than the precision, which means the model usually causes miss alarm alerts. Regarding the self-supervised learning method, the F1 score of the model is low, particularly only 0.8861 for the Fuzzy attack. Although the labeled data of RPM is included in the training set, the model achieved the F1 score of only 0.9850, lower by approximately 1\% compared to our method. Thus, we can conclude that our model is more stable compared to the other models because it has F1 scores higher than 0.99 for all kinds of attacks. 


\subsection{Model complexity analysis}

This section presents model complexity analysis in terms of the number of parameters and inference time. We also compared our model with the DCNN model \cite{Song2020In-vehicleNetwork} to demonstrate that our model is lightweight and fast. The number of parameters directly affects the training and testing time of a model. In theory, the lower the number of parameters, the faster the training and testing model. For the empirical aspect, we also consider the inference time, which is related to the detection latency. The study from \cite{Song2020In-vehicleNetwork} indicated that a small batch size and optimizing inference time reduce the overall detection latency. We set the batch size to one frame and measure the inference time of our model. It is noticeable that the number of parameters of our model includes those of encoder, decoder, and two discriminators, whereas only encoder is used for measuring the inference time.
As illustrated in Table \ref{tab:cmp_dcnn_time}, the total number of parameters of the proposed model is 2.15 million, which is nearly one-fifth of those of the DCNN. Moreover,
we decrease the inference time by approximately eight times with GPU processing and ten times with CPU processing.

\section{Conclusion \label{sec:conclusion}}

The aim of this research is to develop a lightweight and efficient deep learning model for the in-vehicle IDS, using less labeled data as much as possible. We proposed a CAAE model, which is a semi-supervised learning-based in-vehicle IDS to detect CAN bus message injection. The training process consists of three phases, including reconstruction, regularization, and supervised learning phases. With the main idea of utilizing a large amount of unlabeled data, the model learns manifolds of both normal and attack patterns. Therefore, the model provides good results with only a small number of labeled data.

We conducted various experiments to illustrate the performance of the proposed model. A real-world car dataset with four types of message injection attacks was utilized for all the experiments. The results show that feeding only 40\% labeled CAN ID frames to the model achieves the lowest ER of 0.1\% with an F1 score of 0.9984, compared to other supervised methods. In addition, the model can detect unknown attacks, which are not labeled during training, with an F1 score of approximately 0.98 on average. Moreover, the proposed model consists of approximately 2 million trainable parameters, which is very small compared to other deep learning models. Therefore, the proposed model can detect a new attack sample within a millisecond. 

We believe that our model can be helpful in reducing the effort required for labeling and collecting attack data, which is time-consuming in real life. Our model is suitable when there is a large number of unlabeled data but only a small number of those are labeled. In addition, we can collect real car data from users. With the proposed scheme, the model can detect stealthy attacks if they exist in the data collection without requiring a labeling process. 
However, the research only focused on message injection attacks, there are also other kinds of attacks, such as replay or drop attacks. Our future work will include investigating other kinds of attacks in the CAN bus system and applying the proposed model to these data.



\printcredits

\bibliographystyle{elsarticle-num}

\bibliography{references.bib}

\bio{}
\endbio


\end{document}